\documentclass[twocolumn,prd,nofootinbib,showpacs,preprintnumbers]{revtex4}
\usepackage{epsfig}
\usepackage{color}

\newcommand{\be}{\begin{equation}}
\newcommand{\ee}{\end{equation}}

\begin{document}
\preprint{LAPTH-024/12}
\title{Spectral breaks as a signature of cosmic ray induced turbulence in the Galaxy}
\author{Pasquale Blasi$\,{}^{e,p}$, Elena Amato$\,{}^{e}$ and Pasquale D.~Serpico$\,{}^{\bar p}$}
\affiliation{$^{e}$INAF-Osservatorio Astrofisico di Arcetri, Largo E. Fermi, 5 50125 Firenze, Italy}
\affiliation{$^{p}$INFN-Laboratori Nazionali del Gran Sasso, Assergi, Italy}
\affiliation{$^{\bar p}$LAPTh, Univ. de Savoie, CNRS, B.P.110, Annecy-le-Vieux F-74941, France}

\date{\today}

\begin{abstract}
We show that the complex shape of the cosmic ray (CR) spectrum, as recently measured by PAMELA and inferred from Fermi-LAT $\gamma$-ray observations of  molecular clouds in the Gould belt, can be naturally understood in terms of basic plasma astrophysics phenomena. A break from a harder to a softer spectrum at blue rigidity $R\simeq 10$ GV follows from a transition from transport dominated by advection of particles with Alfv\'en waves to a regime where diffusion in the turbulence generated by the same CRs is dominant. A second break at $R\simeq 200$ GV happens when the diffusive propagation is no longer determined by the self-generated turbulence, but rather by the cascading of externally generated turbulence (for instance due to supernova (SN) bubbles)  from large spatial scales to smaller scales where CRs can resonate. Implications of this scenario for the cosmic ray spectrum, grammage and anisotropy are discussed. 
\end{abstract}

\maketitle
{\it Introduction}---The spectrum of cosmic rays (CRs) as observed at the Earth is not a perfect power law, even below the knee, at $\sim 10^{15}$ eV. Recent measurements by the PAMELA experiment have shown that the spectrum has a change of slope at blue $\sim 230$ GV, e.g., for protons, from $\propto E^{-2.85}$ for $E<230$ GeV to blue $\propto E^{-2.67}$ for $E>230$ GeV \cite{pamela}. Moreover, Fermi-LAT $\gamma$-ray data between 100 MeV and 100 GeV from molecular clouds in the Gould belt (but off the Galactic disc) show evidence for a composite CR spectrum, with a slope $\alpha\approx 1.9$ below $\sim 10$ GeV and $\alpha\approx 2.9$ between 10 and 200 GeV \cite{nero}. This analysis has recently been repeated by \cite{kach} who find very similar results. The spectrum of CR protons as measured by PAMELA in the highest energy bins is compatible with balloon measurements by CREAM \cite{cream}. At energies below a few tens of GeV the spectrum observed by PAMELA is  affected by solar modulation and hence it is non-trivial to use these data for comparison with the findings of \cite{nero}.

These spectral features are so far either attributed to propagation effects from nearby sources or modeled with corresponding breaks in either the injection spectrum of CRs, or in the diffusion coefficient experienced by these particles on their way through the Galaxy \cite{moska}. Since the CR flux at Earth is dominated by distant sources, it appears rather unlikely that nearby sources may produce fluctuations of order unity at these energies \cite{amato1}. Recently \cite{tomassetti} proposed that the change of spectrum may be induced by a diffusion coefficient that is not separable in energy and space.

In this Letter we propose that the complex observed CR spectrum is telling us something about how and why particles diffuse in the interstellar medium (ISM) and derives from diffusion of CRs on a background of waves partly due to self-generation and partly to wave-wave turbulent cascading from a large scale, probably corresponding to the size of SN bubbles. 

Several authors have discussed the possibility that CRs could be self-confined by the waves generated through the streaming instability that they excite in the direction of their spatial gradient (see \cite{cesarsky,wentzel} for reviews). In particular \cite{skilling} and \cite{holmes} discussed the effect of self-generation in the presence of ion-neutral damping and non-linear Landau damping (NLLD). The general conclusion in both cases is rather interesting: in the Galactic disk and its vicinities, waves are damped so fast that CR transport is ballistic, while particles are dragged at the Alfv\'en speed (confinement) only in the halo, much above or below the Galactic disc, where the neutral gas density is sufficiently low to avoid effective ion-neutral damping. However, the results of these early works were not based on a self-consistent solution of the transport equation.
More recently, Ref.~\cite{plesser} considered the problem of propagation of CRs on their way out of the sources (e.g. SNRs): the authors find that bubbles of self-generated waves develop in few hundred pc around SNRs, and that such waves dominate the diffusive CR propagation. The effective diffusion coefficient is found to be $D(E)\propto E^{0.6}$. This slope is roughly consistent with the slope of the Boron/Carbon ratio, but leads to severe problems with CR anisotropy if extrapolated to $\gtrsim$ TeV energies \cite{ptuskin,amato2}. Moreover such approaches cannot explain the complex CR spectrum discussed above since the dependence of $D(E)$ on energy does not change. 

Our current knowledge of the Galaxy suggests that, although the average density of neutral gas is relatively high ($\sim {\rm 1\ cm}^{-3}$), such gas is actually confined to regions with small volume filling factor \cite{filling}. Most of the Galaxy is filled with a tenuous ionized gas, where ion-neutral damping is slow and the wave dynamics is most likely determined by NLLD.  The implications of the resonant absorption of MHD waves by CRs in the Galaxy have recently been investigated in \cite{galprop} using GALPROP. 

Moreover we have evidence that turbulence exists in the ISM with a roughly Kolmogorov-type spectrum up to spatial scales of order $\sim 50$ pc or even larger \cite{kolmo}. This turbulence is probably injected by SN explosions on such scales and then cascades towards smaller scales. It only becomes effective for CR scattering when the wavelength becomes as small as the particle Larmor radius. Diffusion models routinely used in calculations of CR propagation are inspired (implicitly or explicitly) by the assumption that something like this happens. 

In this Letter we present our calculations of the combined effect on CR scattering of turbulence cascading from some large scale ($L_0=50\,$pc) through NLLD and self-generated waves induced by CR streaming in the Galaxy. We find that a change in the scattering properties of the ISM must occur at $\sim  200-300$ GV reflecting in a change of shape of the CR spectrum at the same rigidity. While the transition energy can be estimated analytically, we solve the full system of equations describing CR transport and wave evolution so as to obtain a self-consistent spectrum of CRs. In this way, we also find that at energies below $\sim 10$ GV the advection of CRs with waves moving with the Alfv\'en velocity leads to a spectral hardening. Both spectral features are observed \cite{nero,kach,pamela}, and this work was actually stimulated by these observations.

{\it The calculation}---We solve the CR diffusion equation
\be
-\frac{\partial}{\partial z}\left[ D \frac{\partial f}{\partial z}\right] + v_{\rm A}\frac{\partial f}{\partial z} - \frac{dv_{\rm A}}{dz}\frac{p}{3} \frac{\partial f}{\partial p}= q_{\rm CR}(z,p)
\label{eq:transport}
\ee
coupled with the equation for the waves:
\be
\frac{\partial}{\partial k}\left[ D_{kk} \frac{\partial W}{\partial k}\right] + \Gamma_{\rm CR}W = q_{W}(k). 
\label{eq:cascade}
\ee
Here $f(p,z)$ is normalized so that the number of particles in the range $dp$ around momentum $p$ at the location $z$ is $4\pi p^{2} f(p,z) dp$. The diffusion coefficient is related to the wave spectrum through the well known expression~\cite{skilling}:
\be
D(p) = \frac{1}{3} r_{L}(p) v(p) \frac{1}{k\ W(k)},
\label{eq:diff}
\ee
where the power in the form of waves, $W(k)$, satisfies:
\be 
\int_{k_{0}}^{\infty} dk\ W(k) = \eta_B =\frac{\delta B^{2}}{B_{0}^{2}},
\ee
with $\delta B^{2}/4\pi$ the power in turbulent fields and $B_{0}$ the regular magnetic field strength. In Eq.~(\ref{eq:diff}) the momentum and wavenumber are related through the resonance condition $k=1/r_{L}(p)=q B_{0}/(p c)$, with $r_{L}$ the Larmor radius of particles with momentum $p$ moving in the magnetic field $B_{0}$. The underlying assumption is that $\delta B \ll B_0$. In Eq.~(\ref{eq:transport}) we use a simple injection model in which all CRs are produced by SNRs in an infinitely thin disc of radius $R_{d}$: 
\be
q_{\rm CR}(p,z)=\frac{\xi_{\rm CR} E_{\rm SN}{\cal R}_{\rm SN}}{\pi R_{d}^{2} {\cal I}(\alpha) c (m c)^{4}} \left(\frac{p}{m c}\right)^{-\alpha} ~ \delta(z) \equiv q_{0}(p)\delta(z).
\ee
Here $\xi_{\rm CR}$ is the fraction of the total kinetic energy of a SN, $E_{\rm SN}$, assumed to be channelled into CRs, and the SN rate is ${\cal R}_{\rm SN}$. The quantity ${\cal I}(\alpha)=4\pi \int_{0}^{\infty} dx~x^{2-\alpha} \left[ \sqrt{x^{2}+1}-1\right]$ comes from the normalization of the kinetic energy of the SN that goes into CRs. Notice that the particle spectrum is assumed to be a power law in momentum, as expected for diffusive shock acceleration in the test particle regime. 

Eq.~(\ref{eq:cascade}) describes the stationary wave spectrum $W(k)$ under the effect of wave-wave coupling and amplification of waves due to streaming instability at a rate $\Gamma_{\rm cr}(k)$. The cascade is due to NLLD and is described as a diffusion process in $k$-space with a diffusion coefficient \cite{bill}:
\be
D_{kk} = C_{\rm K} v_{\rm A} k^{7/2} W(k)^{1/2}
\label{eq:dkk}
\ee
for a Kolmogorov phenomenology ($C_{\rm K}\approx 5.2\times 10^{-2}$ \cite{zira}). One can easily check that, in the absence of a CR-induced contribution, this diffusive process in $k$-space leads to the standard Kolmogorov spectrum $W(k)\propto k^{-5/3}$ (for $k\gg k_{0}$), if the injection of power occurs at a single $k_{0}=1/L_{0}$. The effect of CRs is to amplify the waves through streaming instability, with the growth rate \cite{skilling}:
\be
\Gamma_{\rm cr}(k)=\frac{16 \pi^{2}}{3} \frac{v_{\rm A}}{k\,W(k) B_{0}^{2}} \left[ p^{4} v(p) \frac{\partial f}{\partial z}\right]_{p=q B_{0}/kc}\ 
\label{eq:gammacr}
\ee 
where the spatial gradient of CRs can be found by solving the transport equation, Eq.~(\ref{eq:transport}).

Eq.~(\ref{eq:transport}) is solved in the simplifying assumptions that $D$ depends weakly on the the $z$-coordinate, and that the Alfv\'en speed is also independent of $z$, except for the fact that Alfv\'en waves move upward (downward) above (below) the disk. This implies that $dv_{\rm A}/dz = 2v_{\rm A}\delta(z)$. With these assumptions the solution of Eq.~(\ref{eq:transport}) can readily be found to be in the form:
\be
f(z,p) = f_{0}(p) \frac{1-e^{-\zeta(1-|z|/H)}}{1-e^{-\zeta}}\,,\:\:\:\:\:\: \zeta(p)\equiv\frac{v_{\rm A} H}{D(p)}\, ,
\label{eq:solution}
\ee
and $f_{0}(p)$ has to satisfy the following equation, obtained by integrating Eq.~(\ref{eq:transport}) in the range $z=(0^{-}-0^{+})$:
\be
-2 D(p) \left[\frac{\partial f}{\partial z}\right]_{z=0^{+}} - \frac{2}{3} v_{\rm A} p \frac{df_{0}}{dp} = q_{0}(p).
\ee
The space derivative can be easily derived from Eq.~(\ref{eq:solution}):
\be
\left[\frac{\partial f}{\partial z}\right]_{z=0^{+}} = \frac{v_{\rm A} f_{0}}{D(p)} \frac{1}{\lambda(p)},~~~~~\lambda(p)=1-\exp\left[\zeta(p)\right].
\ee
Solving for $f_{0}$, the CR spectrum in the disc of the Galaxy is readily found to be:
\be
f_{0}(p)=\frac{3}{2 v_{\rm A}} \int_{p}^{\infty} \frac{dp'}{p'}q_{0}(p) \exp\left[ \int_{p}^{p'}\frac{dp''}{p''} \frac{3}{\lambda(p'')}\right]\ .
\label{eq:f0p}
\ee
In the high energy limit, where diffusion prevails upon convection at speed $v_{\rm A}$, Eq.~(\ref{eq:f0p}) reduces to the well known solution of the diffusion equation in one dimension, $f_{0}^{\rm diff}(p)=q_{0}(p)H/(2 D(p))$.

Eqs.~(\ref{eq:transport}) and (\ref{eq:cascade}) form a set of two non-linear differential equations. We solve them iteratively so that the final results are the spectrum of CRs in the Galactic disc, $f_{0}(p)$, and the power spectrum of waves, $W(k)$, resulting from self-generation and cascading. Before illustrating the exact results, it is useful to estimate the energy of CR protons where one expects a transition from self-generated waves to waves deriving from Kolmogorov cascade. 
In order to do this, we use the fact that CR driven waves saturate when the NLLD rate ($\Gamma_{NL}\approx D_{kk}/k^2$, with $D_{kk}$ given in Eq.~(\ref{eq:dkk})) equals the growth rate $\Gamma_{\rm cr}$ (Eq.~(\ref{eq:gammacr})). Using the expression
$f_0(p)=A_p (p/mc)^{-\gamma_p}$ to describe the spectrum, with $A_p$ and $\gamma_p$ taken from PAMELA data above 250 GeV, we find:
\be
W_{\rm CR}=\left[ \frac{16 \pi^2}{3C_K}\frac{A_p (mc)^4}{B_0^2H}\left( \frac{eB_0}{mc^2}\right)^{4-\gamma_p}\right]^\frac{2}{3}\ k^{\frac{2}{3}(\gamma_p-\frac{13}{2})}, 
\label{eq:wcr}
\ee 
where $k$ is related to the particle energy through the resonance condition.
The transition length-scale, and hence energy, is then found by simply equating $W_{\rm CR}$ to the externally generated Kolmogorov spectrum. 
We take the latter in the form: 
\be
W_{\rm ext}(k)=(2\eta_B/3k_0)\ (k/k_0)^{-5/3}\ .
\label{eq:wext}
\ee
Rather than guessing the value of $\eta_B$, we use again our knowledge of the CR spectrum based on Pamela data, and express $\eta_B$ in terms of the CR acceleration efficiency $\xi_{\rm CR}$. This can be done by just recalling that in the high energy regime $f_0(p)$ is well approximated by $f_0^{\rm diff}(p)$, with $D(p)$ given by Eq.~(\ref{eq:diff}). At scales at which the Kolmogorov turbulence is the dominant scattering source, one finds:   
\be
\eta_B=\frac{\pi R_d^2 I(\alpha) A_p \left(m c^2\right)^3}{c H \xi_{\rm CR} E_{\rm SN} {\cal R}_{\rm SN}}\left( \frac{mc^2}{q B_0}\right)^{1/3}\ k_0^{2/3}\ .
\label{eq:etaB}
\ee
Using this condition in Eq.~(\ref{eq:wext}) and equating it to Eq.~(\ref{eq:wcr}) we obtain:
\be
E_{\rm tr}=228\ {\rm GeV} \left(\frac{R_{d,10}^2 H_{3}^{-1/3}}{\xi_{0.1}E_{51}{\cal R}_{30}}\right)^{\frac{3}{2(\gamma_p-4)}} B_{0,\mu}^{\frac{2\gamma_p-5}{2(\gamma_p-4)}},
\label{eq:etr}
\ee
where $R_{d,10}=R_{d}/10\,\rm kpc$, $H_{3}=H/3\,\rm kpc$, $\xi_{0.1}=\xi_{\rm CR}/0.1$, $E_{51}=E_{\rm SN}/10^{51}\rm erg$, ${\cal R}_{30}={\cal R}_{\rm SN}/30\, {\rm yr}^{-1}$, $B_{0,\mu}=B_{0}/\mu$G.

The estimate obtained for the reference values of the parameters is tantalizingly close to the energy where PAMELA data show a change of slope of the spectrum of protons from $E^{-2.85\pm 0.015}$ ($E< 230$ GeV) to $E^{-2.67\pm 0.03}$ ($E> 230$ GeV). Clearly the fact that al low $k$'s (large momenta) the power spectrum is $W(k)\sim k^{-1/3}$ implies that the CR injection spectrum must be $q_{0}(p)\sim p^{-4.3}$. We notice that the transition energy turns out to be independent of the characteristic scale of turbulence, $1/k_0$: this fact is especially important given the large uncertainty on this parameter. What does depend on $k_0$, linearly, is the energy density in turbulent magnetic field, which for the above values of the parameters and  $L_0=50\,{\rm pc}$ turns out to correspond to $\xi_B\approx$ 8\%.

Interestingly enough, taking into account that in the advection dominated regime the solution is $f_{0}^{\rm adv}(p)\approx q_{0}(p)/v_{\rm A}$ and equating this to the solution in the diffusion regime determined by self-generated waves, it is easy to see that for reasonable values of the Alfv\'en speed the low energy transition to a spectrum $f_{0}(p)\propto p^{-\alpha}$ occurs at $E\lesssim 10$ GeV.

\begin{figure}[htb!!!!!!]
\vspace{-1.5pc}
\begin{center}
\begin{tabular}{c}
\epsfig{figure=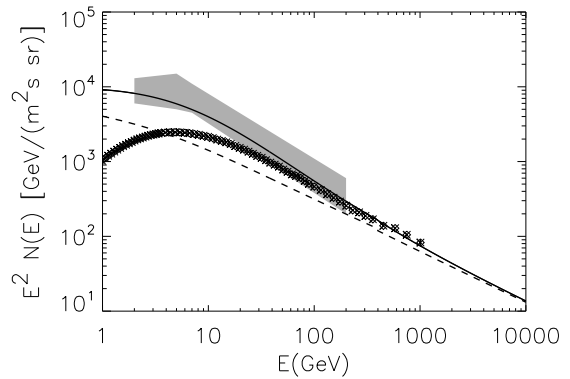,width=0.92\columnwidth}
\end{tabular}
\end{center}
\vspace{-2pc} \caption{Spectrum of CR protons from our calculations (solid line) compared with the results from observations of $\gamma$-ray emission from clouds in the Gould belt \cite{nero} (shaded region) and with PAMELA measurements \cite{pamela}.}\label{fig:results}
\end{figure}

{\it Results and Discussion}--- The iterative procedure described above leads to the spectrum of CRs plotted in Fig.~\ref{fig:results} (solid line). 
The symbols are PAMELA data. Since solar modulation affects the low energy part, only the data above 80 GeV were used in \cite{pamela} to highlight the break at $\sim 230$ GV. The dashed line is the solution of the same equations but setting the self-generation term to zero. The shaded area shows the CR spectrum inferred in Ref. \cite{nero}. 
\begin{figure}[htb!!!!!]
\vspace{-1.5pc}
\begin{center}
\begin{tabular}{c}
\epsfig{figure=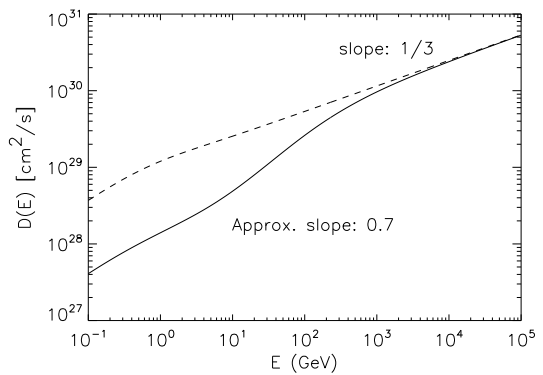,width=0.88\columnwidth}
\end{tabular}
\end{center}
\vspace{-2pc} \caption{Diffusion coefficient induced by streaming instability of CRs and cascading from a large spatial scale of 50 pc.}\label{fig:diff}
\end{figure}
The break at $\sim 200$ GeV reflects the transition from a regime where the scattering centers are self-generated to a regime where particles diffuse in external turbulence that cascades from larger spatial scales. This is well visible in Fig.~\ref{fig:diff}, where we plot the diffusion coefficient with and without self-generation (solid and dashed line respectively). The energy dependence of the diffusion coefficient in the energy range $10<E<200$ GeV is not a perfect power law, but if we approximate it as such, we find $D(E)\sim E^{0.7}$. At larger energies, $E>200$ GeV, the trend is $D(E)\sim E^{1/3}$, as expected for a Kolmogorov cascade. At low energy the change of slope is due to the transition to a non-relativistic regime, but there the propagation is advection-dominated.

\begin{figure}[ht!!!!!]
\vspace{-1.5pc}
\begin{center}
\begin{tabular}{c}
\epsfig{figure=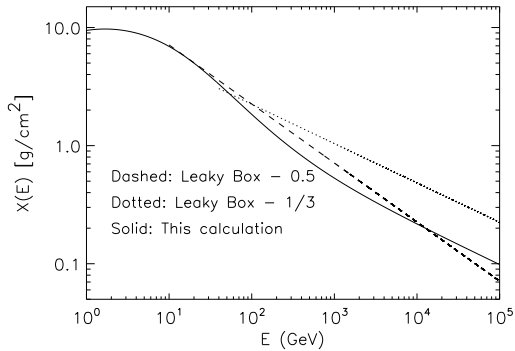,width=0.84\columnwidth}
\end{tabular}
\end{center}
\vspace{-2pc} \caption{Grammage obtained in our calculations (solid line) compared with the leaky box fit proposed in Ref.~\cite{leaky} (dashed and dotted lines are the fit for $\delta=0.5$ and 1/3 respectively).}\label{fig:gram}
\end{figure}
The grammage traversed by CRs of velocity $v$ under the combined effect of diffusion and advection at the Alfv\'en speed is easily found to be
$X(E) = \frac{\mu\, v}{2\,v_{\rm A}}\left[ 1 - \exp\left(-\frac{v_{\rm A}H}{D(E)}\right)\right],$
where $\mu=2\, n\, m\, h = 2.4 {\rm mg/cm}^{2}$ is the surface density of the disc (corresponding to a Galactic disc density $n_{d}\simeq 1~{\rm cm}^{-3}$, for a typical chemical composition of the interstellar matter). The grammage obtained in our calculations is shown in Fig.~\ref{fig:gram} (solid line) and compared with the leaky box fits proposed in \cite{leaky} (dashed (dotted) line for $\delta=0.5$ ($\delta=1/3$)). The normalization in the $1-10$ GeV range comes out naturally in our calculations, together with the change of slope in the diffusion properties at $\approx 200\,$GeV.

The flattening in the energy dependence of the diffusion coefficient at high energies automatically avoids severe problems with CR anisotropy. The mean anisotropy amplitude is $\delta \sim 10^{-3}$ at 1 TeV and increases with energy as $E^{1/3}$. However, as discussed in \cite{amato2}, this mean value is not very meaningful in that the amplitude is dominated by the nearest and most recent sources, and may dramatically differ from the mean value. Nevertheless, it is encouraging that the physical processes described here naturally lead to steep energy dependence of the diffusion coefficient in the low energy regime that does not necessarily lead to violate observed data on anisotropy. 

In summary, we showed that in a simplified but physically consistent model for the CR propagation in the ISM the departure from a power-law spectrum in CRs measured at Earth is a consequence of basic processes. Both a convective velocity of the order of $v_{\rm A}$ in Eq.~(\ref{eq:transport}) and the contribution of CRs to the wave spectrum in Eq.~(\ref{eq:cascade}) are unavoidable, albeit often neglected in phenomenological studies.  
It is actually remarkable that without ad hoc free parameters, the basic trend shown by the measured and inferred CR spectra can be reproduced in a relatively simple model.
The several approximations that have been made (e.g. treatment of non-linear damping, absence of re-acceleration, neglect of non-linear effects in diffusive shock acceleration) are not expected to change the qualitative picture that emerges.
The scenario detailed here also allows to accommodate the inferred behaviour of the grammage as well as---at least qualitatively---the stringent constraints coming from anisotropy at high energy. We find encouraging that the recent {\it bonanza} of CR data allows one to gain some insights on (astro)physical processes involving CRs, beyond the mere task of providing more accurate {\it fits} to injection spectra or propagation parameters; this is a trend which hopefully will be further boosted in the near future with the expected results of AMS-02.

{\it Acknowledgements---} We thank A. Neronov, A. Taylor for discussions and V. Formato and M. Boezio for providing PAMELA data points.
 PB thanks LAPTh for hospitality during the initial phases of this project. The work of PB and EA is partially funded through PRIN 2010 and ASTRI grants. 


\begin{thebibliography}{00}

\bibitem{pamela}
O., Adriani {\it et al.}, Science, {\bf 332}, 69 (2011). 

\bibitem{amato1}
P. Blasi, and E. Amato, JCAP {1}, 010 (2012).

\bibitem{nero}
A. Neronov, D.V. Semikoz, and A.M. Taylor, PRL {\bf 108} 1105 (2012).

\bibitem{kach}
M. Kachelrie{\ss}, S. Ostapchenko, arXiv:1206.4705.

\bibitem{cream}
H.S. Ahn, et al., ApJ Lett. {\bf 714}, 89 (2010). 

\bibitem{moska}
A.E. Vladimirov {\it et al.},  ApJ, {\bf 752}, 68 (2012).

\bibitem{tomassetti}
N. Tomassetti, ApJ Lett., {\bf 752}, 13 (2012).

\bibitem{cesarsky}
C.J. Cesarsky, Ann. Rev. of A\&A, {\bf 18}, 289 (1980).

\bibitem{wentzel}
D.G. Wentzel, Ann. Rev. of A\&A, {\bf 12}, 71 (1974).

\bibitem{skilling}
J. Skilling, ApJ ,{\bf 170}, 265 (1971).

\bibitem{holmes}
J.A. Holmes, MNRAS, {\bf 170}, 251 (1975).

\bibitem{plesser}
V.S. Ptuskin, V.N. Zirakashvili, A.A. Plesser, Adv. Space Res. {\bf 42}, 486 (2008).

\bibitem{ptuskin}
V.S. Ptuskin, F.C. Jones, E.S. Seo, R. Sina, Adv. Space Res. {\bf 37}, 1909 (2006).

\bibitem{amato2}
P. Blasi, and E. Amato, JCAP {1}, 011 (2012).

\bibitem{filling}
M.A. Dopita \& R.S. Sutherland 2002,  Astrophysics of the Diffuse Universe (Berlin: Springer-Verlag). 

\bibitem{galprop}
V. Ptuskin, et al., ApJ, {\bf 642}, 902 (2006).

\bibitem{kolmo}
J.W. Armstrong, B.J. Rickett, S.R. Spangler, ApJ, {\bf 443}, 209 (1995).

\bibitem{zira}
V.S. Ptuskin, V. N. Zirakashvili, A\&A {\bf 403} 1 (2003).

\bibitem{leaky}
V.S. Ptuskin, O.N. Strelnikova, L.G. Sveshnikova, Astrop. Phys. {\bf 31} 284 (2009).

\bibitem{bill}
Y. Zhou, W.H. Matthaeus, J. of Geophys. Res., {\bf 95}, 14881 (1990).

\end{thebibliography}
\end{document}